\documentclass[a4paper,11pt]{article}
\usepackage{jinstpub} 

\usepackage{siunitx}
\usepackage{upgreek}
\usepackage{comment}
\usepackage{url,hyperref}
\newcommand{\authorlist}{

\author[1]{V. Boccia}
\author[2]{T. Asada}
\author[1]{A. Alexandrov}
\author[3]{G. Ambrosi}
\author[4,5]{S. Argirò}
\author[3]{M. Barbanera}
\author[6,5]{N. Bartosik}
\author[7]{G. Battistoni}
\author[8]{A. Bigot}
\author[9,10]{M. G. Bisogni}
\author[11]{G. Butella}
\author[5]{F. Cavanna}
\author[5]{P. Cerello}
\author[9,10]{E. Ciarrocchi}
\author[12]{N. D'Ambrosio}
\author[13,1]{G. De Lellis}
\author[13,1]{A. Di Crescenzo}
\author[14,15]{B. Di Ruzza}
\author[16,17]{M. Dondi}
\author[11,5]{M. Donetti}
\author[22]{Y. Dong}
\author[13,18]{M. Durante}
\author[19,20]{R. Faccini}
\author[4,5]{V. Ferrero}
\author[8]{C. Finck}
\author[5]{E. Fiorina}
\author[1]{M. Francesconi}
\author[16,17]{M. Franchini}
\author[21,20]{G. Franciosini}
\author[10]{L. Galli}
\author[13]{A. Iuliano}
\author[3]{K. Kanxheri}
\author[5]{B. Kharpuse}
\author[23,24]{M. Kimura}
\author[25]{S. Kodaira}
\author[10]{A. C. Kraan}
\author[13,1]{A. Lauria}
\author[26,4]{E. Lopez Torres}
\author[27]{T. Maggipinto}
\author[21,20]{M. Magi}
\author[16,17]{A. Manna}
\author[28,20]{M. Marafini}
\author[12]{S. Masci}
\author[10]{M. Massa}
\author[16,17]{C. Massimi}
\author[7]{I. Mattei}
\author[29,3]{S. Mazzolani}
\author[16]{A. Mengarelli}
\author[11]{A. Mereghetti}
\author[19,20]{R. Mirabelli}
\author[10]{A. Moggi}
\author[30,1]{M. C. Montesi}
\author[31,32]{M. C. Morone}
\author[9,10]{M. Morrocchi}
\author[7]{S. Muraro}
\author[2,36]{T. Naka}
\author[5]{N. Pastrone}
\author[21,20]{V. Patera}
\author[5]{F. Pennazio}
\author[17,16]{C. Pisanti}
\author[33,3]{P. Placidi}
\author[11]{M. Pullia}
\author[19,20]{F. Quattrini}
\author[16]{S. Rabaglia}
\author[6,5]{L. Ramello}
\author[18]{C. A. Reidel}
\author[16]{R. Ridolfi}
\author[31,32,28]{L. Rocchetti}
\author[34]{L. Sabatini}
\author[33,3]{L. Salvi}
\author[34]{C. Sanelli}
\author[21,20]{A. Sarti}
\author[36]{O. Sato}
\author[11]{S. Savazzi}
\author[21,20]{A. Schiavi}
\author[18]{C. Schuy}
\author[35]{E. Scifoni}
\author[3]{L. Servoli}
\author[3]{G. Silvestre}
\author[6,5]{M. Sitta}
\author[2]{K. Someya}
\author[4,5]{B. Spadavecchia}
\author[16]{R. Spighi}
\author[34]{E. Spiriti}
\author[19,20,28]{L. Testa}
\author[1]{V. Tioukov}
\author[34]{S. Tomassini}
\author[24,35]{F. Tommasino}
\author[21,20]{M. Toppi}
\author[20]{G. Traini}
\author[34]{A. Trigilio}
\author[16,17]{G. Ubaldi}
\author[17,16]{S. Valentinetti}
\author[8]{M. Vanstalle}
\author[17,16]{M. Villa}
\author[18,33]{U. Weber}
\author[17,16]{R. Zarrella}
\author[17,16]{A. Zoccoli}
\author[27,15]{G. Galati}

\affiliation[1]{INFN, Section of Napoli, Napoli, Italy}
\affiliation[2]{Department of Physics, Faculty of Science, Toho University, Chiba, Japan}
\affiliation[3]{INFN, Section of Perugia, Perugia, Italy}
\affiliation[4]{University of Torino, Department of Physics, Torino, Italy}
\affiliation[5]{INFN, Section of Torino, Torino, Italy}
\affiliation[6]{University of Piemonte Orientale, Alessandria, Italy}
\affiliation[7]{INFN, Section of Milano, Milano, Italy}
\affiliation[8]{IPHC, CNRS/Université de Strasbourg, Strasbourg, France}
\affiliation[9]{University of Pisa, Department of Physics, Pisa, Italy}
\affiliation[10]{INFN, Section of Pisa, Pisa, Italy}
\affiliation[11]{CNAO, Pavia, Italy}
\affiliation[12]{INFN, Gran Sasso National Laboratories, Assergi, Italy}
\affiliation[13]{University of Napoli Federico II, Department of Physics E. Pancini, Napoli, Italy}
\affiliation[14]{University of Foggia, Foggia, Italy}
\affiliation[15]{INFN, Section of Bari, Bari, Italy}
\affiliation[16]{INFN, Section of Bologna, Bologna, Italy}
\affiliation[17]{University of Bologna, Department of Physics and Astronomy, Bologna, Italy}
\affiliation[18]{GSI Helmholtzzentrum, Darmstadt, Germany}
\affiliation[19]{Sapienza University of Rome, Department of Physics, Rome, Italy}
\affiliation[20]{INFN, Section of Roma 1, Rome, Italy}
\affiliation[21]{Sapienza University of Rome, SBAI Department, Rome, Italy}
\affiliation[22]{Data Science Unit, Fondazione IRCCS Istituto Nazionale dei Tumori di Milano, Milan, Italy}
\affiliation[23]{Department of Proton Therapy Physics, Nagoya Proton Therapy Center, Nagoya City University West Medical Center, Nagoya, Japan}
\affiliation[24]{University of Trento, Department of Physics, Trento, Italy}
\affiliation[25]{Radiation Measurement Research Group, Institute for Radiological Science, National Institutes for Quantum Science and Technology (QST), Chiba, Japan}
\affiliation[26]{CEADEN, Havana, Cuba}
\affiliation[27]{University of Bari, Department of Physics, Bari, Italy}
\affiliation[28]{Museo Enrico Fermi, Rome, Italy}
\affiliation[29]{University of Camerino, Camerino, Italy}
\affiliation[30]{University of Napoli Federico II, Department of Chemistry, Napoli, Italy}
\affiliation[31]{University of Rome Tor Vergata, Rome, Italy}
\affiliation[32]{INFN, Section of Roma Tor Vergata, Rome, Italy}
\affiliation[33]{University of Perugia, Department of Physics and Geology, Perugia, Italy}
\affiliation[34]{INFN, Frascati National Laboratories, Frascati, Italy}
\affiliation[35]{TIFPA-INFN, Trento, Italy}
\affiliation[36]{Kobayashi-Maskawa Institute, Nagoya University, Nagoya, Japan}

}

\title{\boldmath Toward High-Resolution Detection of Target Fragmentation: TEA-Sensitized NIT for Proton Therapy Applications}

\authorlist

\abstract{Nano Imaging Trackers (NIT) are fine-grained nuclear emulsions capable of tracking charged particles with sub-micrometric spatial resolution. The DAMON (Direct Measurement of Target Fragmentation) experiment recently employed NIT to detect target-fragmentation events relevant to proton therapy in direct kinematics. These measurements showed that the small crystal size, while providing high spatial resolution, limits the reconstruction efficiency for primary proton tracks and high-energy secondary protons. This work reports an enhancement of NIT sensitivity achieved through alternative chemical sensitization methods, namely gold-plus-sulfur (Au-S) and triethanolamine (TEA), together with the use of the GR-1 developer. The detector response was evaluated with gamma-ray, carbon-ion and proton exposures, including clinically relevant proton energies. 
}


\arxivnumber{} 

\begin{document}
\maketitle
\flushbottom


\section{Introduction} 
\label{sec:intro}

Nuclear emulsion films provide the highest spatial resolution among tracking detectors in particle and nuclear physics~\cite{ariga2020nuclear}. They have played a key role in major discoveries and have recently regained attention thanks to advances in automated optical scanning \cite{alexandrov2020super, alexandrov2023super}. 
The intrinsic spatial resolution of a nuclear emulsion is primarily determined by the average distance between crystals, which depends on their size. Smaller crystals provide higher spatial resolution; however, this improvement comes at the expense of reduced detection efficiency for ionizing particles. Nano Imaging Trackers (NIT) are a class of fine-grained nuclear emulsions with AgBr(I) crystal diameters ranging from 20 to 80~nm~\cite{asada2017development}, approximately an order of magnitude smaller than those of conventional emulsions (about 200 nm). The smaller crystal size imposes constraints on the readout: automated optical microscopes have been developed and successfully applied in previous studies~\cite{katsuragawa2017new, boccia2025dark}. Moreover, super-resolution techniques based on localized surface plasmon resonance (LSPR) have been introduced, enabling the detection of tracks with lengths down to approximately 50~nm~\cite{alexandrov2020super, alexandrov2023super}. 
Owing to their spatial resolution and tunable crystal size, NIT are employed in a broad range of applications, including directional dark matter searches, environmental neutron monitoring and nuclear physics. The NEWSdm experiment employs NIT to detect and reconstruct sub-micrometric nuclear recoils induced by weakly interacting massive particles (WIMPs), exploiting directional information to discriminate signal from background~\cite{agafonova2018discovery, agafonova2023directional}. NIT have also enabled measurements of sub-MeV environmental neutron spectra and angular distributions through the detection of low-energy recoil protons~\cite{shiraishi2021development, shiraishi2023environmental}. More recently, NIT have been employed for the measurement of nuclear fragmentation in particle therapy, where both short-ranged target fragments and projectile fragments contribute significantly to the dose deposition. While the excellent spatial resolution enabled the direct observation of target fragments that are difficult to detect with conventional solid-state detectors~\cite{boccia2025dark}, the reduced sensitivity associated with the small crystal size limited the reconstruction efficiency for weakly ionizing particles, such as secondary protons above approximately 15~MeV. As a result, target fragmentation events could only be identified through the presence of highly ionizing secondary fragments. Energetic secondary protons play an important role in proton therapy~\cite{tommasino2015proton,bellinzona2021biological} and the inability to 
reconstruct them reduced the overall detection efficiency. The objective of this work is to improve the sensitivity of NIT while preserving their intrinsic spatial resolution. To this end, alternative chemical sensitization methods (gold-plus-sulfur and triethanolamine) together with a new chemical developer (GR-1) were investigated. Their performance was evaluated through irradiations with gamma rays, proton beams, and carbon-ion beams.

\section{NIT sensitization and development} 

A nuclear emulsion film consists of silver halide crystals, typically AgBr(I), uniformly dispersed within a gelatin matrix. When a charged particle traverses the sensitive layer, it ionizes the silver halide crystals along its path, producing a latent-image composed of small clusters of silver atoms formed by trapped photoelectrons (latent-image centers). During chemical development, these centers are amplified into visible \emph{silver grains}. Nuclear emulsion films are typically produced as thin sensitive layers deposited onto transparent substrates, such as glass or plastic. NIT production relies on the PVA–gelatin mixing method (PGMM), which allows precise control of crystal growth and prevents aggregation during desalting and redispersion~\cite{asada2017development}. After production, the films are exposed to radiation, chemically developed and analyzed by optical microscopy. Due to the ultra-fine size of the AgBr(I) crystals, contaminated dust, airborne particles, and chemical impurities can produce spurious signals that appear as developed silver grains. For this reason, all preparation steps (melting, mixing with sensitizers, pouring onto support substrates and drying) must be performed in a clean-room environment or, at minimum, under a laminar-flow hood equipped with HEPA filtration. Temperature and relative humidity must also be carefully controlled to prevent uneven drying, crystal aggregation and mechanical defects in the gelatin matrix. 
    
\subsection{Halogen Acceptor Sensitization}\label{sec:HAsensitization}

To achieve sufficient sensitivity to ionizing radiation, nuclear emulsions must undergo a sensitization process, either during gel production or immediately before use. For NIT, where extremely low background is essential, a mild but effective sensitization method based on halogen acceptors (HA) has traditionally been adopted. 
Latent image centers formed by trapped photoelectrons can be partially or completely destroyed by rehalogenation processes involving bromine atoms or molecules generated by ionization-induced holes \cite{tani2024dark}. This effect is particularly significant in ultra-fine-grained emulsions, where the small crystal size enhances electron-hole recombination. Halogen acceptors mitigate this effect by capturing bromine species, thereby suppressing rehalogenation and preserving the latent image centers~\cite{kuge2009sensitization}.
In conventional NIT handling, sodium sulfite is used as a halogen acceptor, as it increases sensitivity without significantly raising the fog density~\cite{NakaPhD}. Sensitization can be applied either by mixing the sodium sulfite solution into the melted emulsion prior to pouring, or by immersing the dried film in a sensitizer solution. In this work, the former approach is adopted to ensure consistency when comparing different sensitization methods.

\subsection{Methol-Ascorbic Acid development}

The Metol-Ascorbic Acid (MAA) developer~\cite{james1953psaMAA} is the standard development method for NIT, owing to its ability to suppress background while preserving track visibility. MAA is a purely chemical (non-physical) developer: amplification occurs at latent image centers on the surface of silver halide crystals. 
For NIT applications, the MAA developer is typically operated at a pH of 10.2 and at a low temperature of 5~$^\circ$C, conditions optimized to suppress thermal activation processes that could otherwise lead to the formation of unwanted silver grains~\cite{NakaPhD}.


\section{Improvement of NIT Sensitivity}\label{sec:first_results}
\label{subsec:newsensitization}

To enhance the sensitivity without increasing the crystal size, two alternative chemical sensitization methods were investigated: gold-plus-sulfur (Au-S) and triethanolamine (TEA). Au-S sensitization is a well-established technique for conventional nuclear emulsions, whereas TEA has previously been studied only in fine-grained photographic materials. In this work, both methods are evaluated for NIT with 70~nm crystals and combined with the GR-1 developer \cite{yamamoto2023Reversaldevelopment} to determine their effectiveness in improving the response to ionizing radiation while preserving the detector's intrinsic spatial resolution.


\subsection{Gold-plus-sulfur sensitization}

Gold-plus-sulfur (Au-S) sensitization is one of the most powerful and well-established chemical treatments for silver halide crystals~\cite{tani2000comprehensive}. It combines sulfur sensitization and gold sensitization, resulting in a synergistic enhancement of latent-image formation. Although Au-S sensitization has been widely applied to conventional nuclear emulsions, prior to this work it had not been studied for NIT due to concerns about fog generation. Given the sensitivity limitations observed with HA‑sensitized NIT, Au-S was reconsidered as a potential method for restoring sufficient latent‑image formation in NIT. Sulfur sensitization is initiated by the addition of sodium thiosulfate, which reacts with the crystal surface to form sulfide ion dimers ($\mathrm{S}_2^{2-}$). These small sulfide aggregates act as electron traps, promoting the formation of latent-image centers. However, in the absence of additional stabilization, sulfur sensitization may lead to the growth of larger silver sulfide ($\mathrm{Ag}_2\mathrm{S}$) clusters, which may become developable even without radiation exposure, thereby increasing fog density. Gold sensitization mitigates this issue. As explained in~\cite{tani2000comprehensive}, 
the incorporation of $\mathrm{Au^{3+}}$ ions at the sulfur‑sensitized sites produces shallow electron traps with enhanced catalytic activity. Gold stabilizes the sulfide dimers, suppressing the formation of large $\mathrm{Ag_2S}$ clusters, and improves the efficiency with which small latent‑image silver clusters trigger development. This stabilizing and catalytic action lowers the development threshold, allowing smaller latent image centers to be successfully amplified during chemical development.
    
\subsection{Triethanolamine sensitization}
    
    

Triethanolamine (TEA) sensitization acts by partially dissolving the surface of the \textit{AgBr(I)} crystals, leading to the formation of $\mathrm{AgOH}$, which subsequently decomposes into $\mathrm{Ag}_2\mathrm{O}$ and $\mathrm{H}_2\mathrm{O}$. The resulting silver oxide is then reduced to metallic silver, enriching the development centers and increasing the sensitivity of the crystals~\cite{ryabova2014TEA}. 
Prior to this work, TEA sensitization had already been applied to fine grained emulsions in~\cite{ryabova2014TEA}, where a 0.5\% TEA solution increased the sensitivity by more than three orders of magnitude at $\lambda = 1.06~\si{\micro\meter}$ and by a factor of 16 at $\lambda = 1.32~\si{\micro\meter}$. However, the sensitivity increase to ionizing radiation had not been investigated. 

\subsection{Sensitivity to electrons}\label{subsec:austealngs}

\begin{figure}[h!]
\centering
\includegraphics[width = 0.8\linewidth]{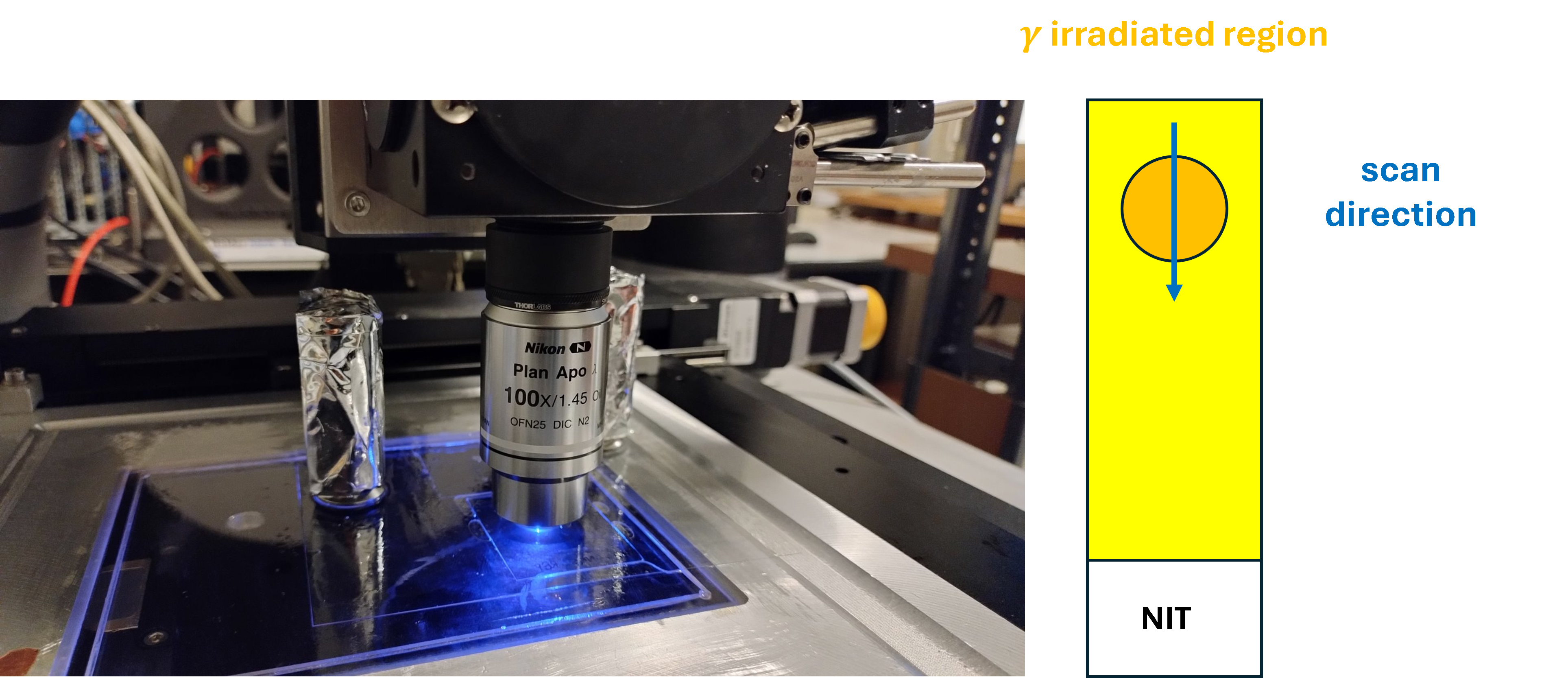}
\caption{Left: scanning of a NIT slide under epi-illumination. Right: schematic of the line scan procedure used to locate the exposure peak and estimate the relative sensitivity increase.}
\label{fig:nit_gamma_scans}
\end{figure}

\begin{figure}[h!]
    \centering
    \begin{minipage}{0.50\linewidth}
        \centering
        \includegraphics[width=\linewidth]{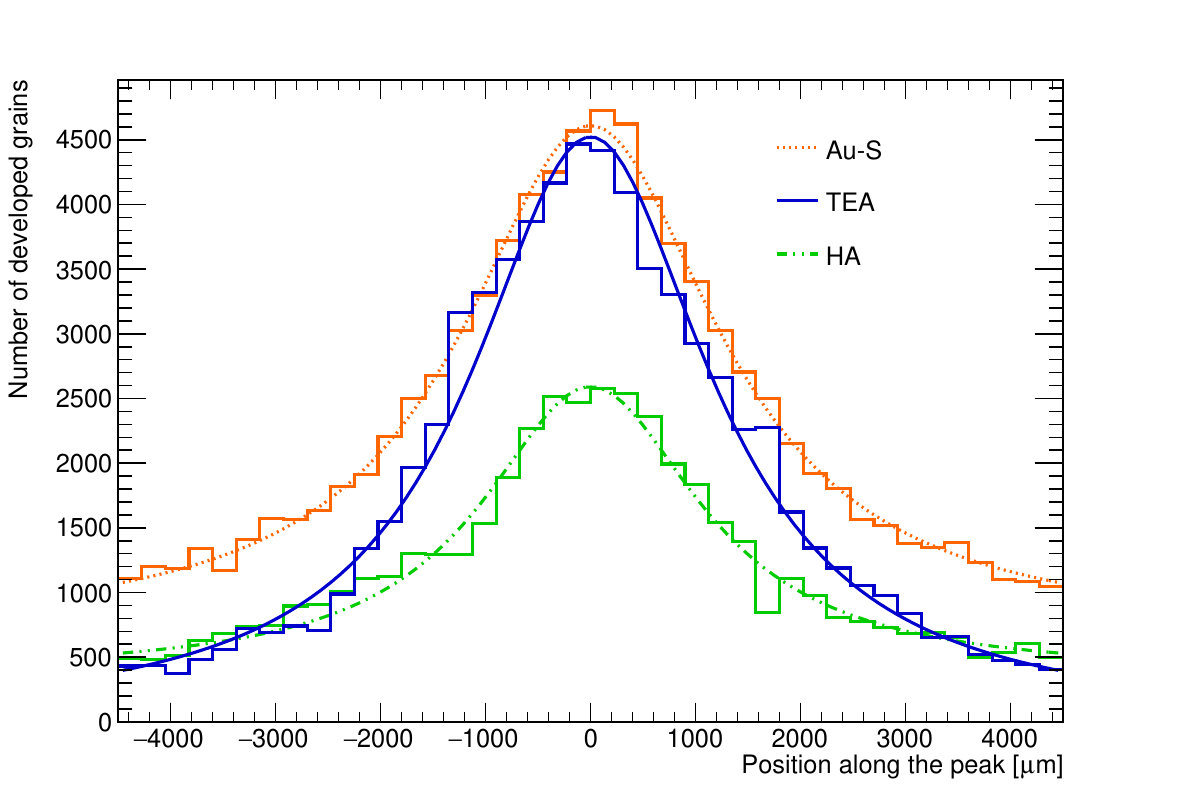}
    \end{minipage}\hfill
    \begin{minipage}{0.50\linewidth}
        \centering
        \includegraphics[width=\linewidth]{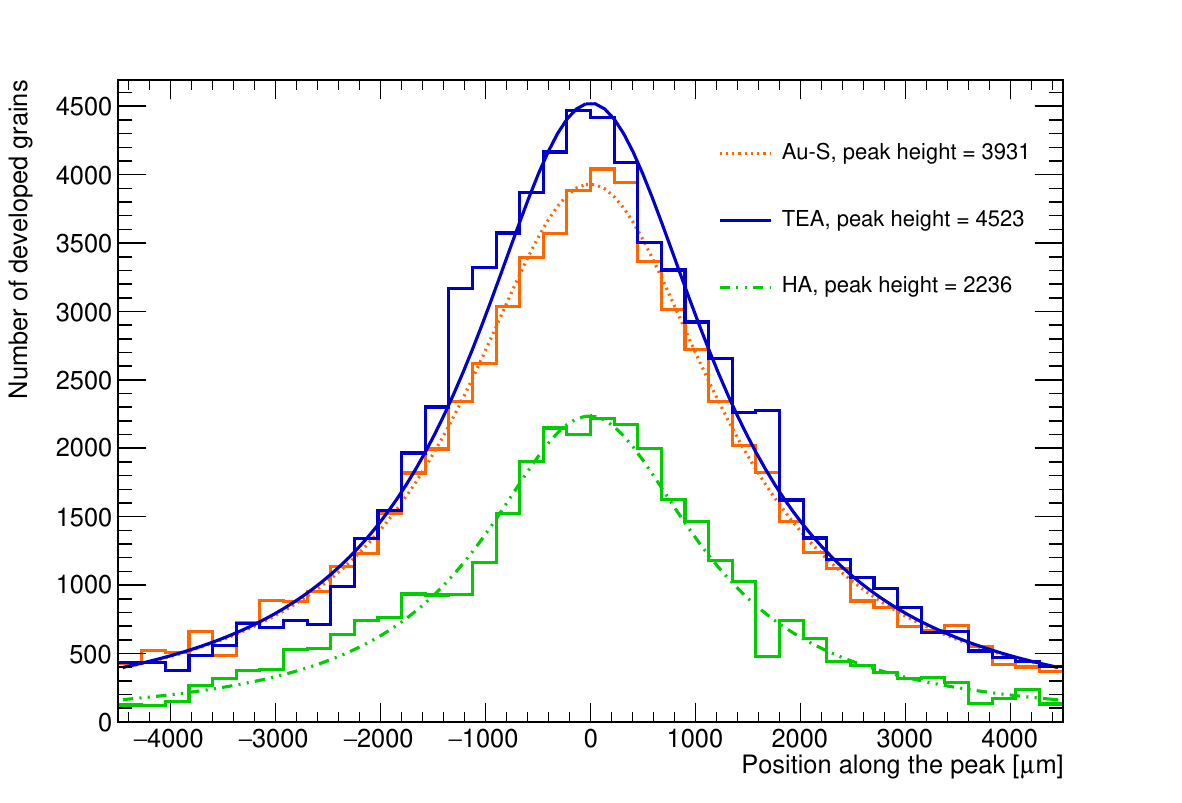}
    \end{minipage}
    \caption{Gamma-ray peak profiles measured with NIT samples prepared with different chemical sensitization methods. Left: reconstructed grain-density profiles showing the increase in developed grains for Au-S and TEA sensitization compared with the HA reference. Right: corresponding profiles after background subtraction, highlighting the enhanced sensitivity of TEA while maintaining a low fog density and improved signal-to-noise ratio.}
    \label{fig:nit_lngs_test_results}
\end{figure}

To compare the performance under the same conditions, NIT samples were prepared with HA, Au-S, and TEA sensitization and then exposed to a $\upgamma$-ray source. For Au-S sensitization, two 0.1\% solutions were prepared: one of sodium thiosulfate pentahydrate ($\mathrm{Na}_2\mathrm{S}_2\mathrm{O}_3 \cdot 5\mathrm{H}_2\mathrm{O}$) and one of gold(III) chloride trihydrate ($\mathrm{HAuCl}_4 \cdot 3\mathrm{H}_2\mathrm{O}$). The emulsion gel was pre-heated to $50~^\circ\mathrm{C}$; the thiosulfate solution was added after 15~min, followed 2~min later by the gold chloride solution. Stirring was continued for 1~h, after which the gel was cooled below $50~^\circ\mathrm{C}$. Two formulations were tested: the standard recipe used for conventional nuclear emulsions and a stronger formulation obtained by scaling the reagent quantities according to the surface-area ratio between conventional 200~nm crystals and 70~nm NIT crystals, $(200/70)^2$. Samples treated with the stronger formulation became visibly darker, indicating over-sensitization, whereas those treated with the standard formulation remained transparent and suitable for analysis.
For TEA sensitization, the solution was added to the molten emulsion gel at about $40~^\circ\mathrm{C}$, with a weight ratio of 17.6\% with respect to the emulsion. Several TEA concentrations, from 5 to 25~g/L, were tested. The sensitized samples were either exposed for 3~min to a $^{241}$Am $\gamma$-ray source (about 60 keV) or directly developed to evaluate the fog density, since an increase in sensitivity is often accompanied by an increase in background.
A visual inspection of the Au-S and TEA-sensitized samples revealed short tracks composed of 2-3 grains, attributed to photoelectrons produced by $\gamma$-ray interactions. This was the first clear indication of an increase in sensitivity with respect to HA-sensitized NIT, which under the same exposure conditions showed only isolated single grains. To quantify this effect, linear scans were performed across the $\gamma$-source peak, as illustrated in Fig.~\ref{fig:nit_gamma_scans}, and the relative sensitivity was inferred from the number of reconstructed grains. 
The peak profiles shown in Fig.~\ref{fig:nit_lngs_test_results} indicate that both Au-S and TEA increase the number of developed grains with respect to HA sensitization. 
Au-S also produced a higher fog density, as indicated by the elevated background level. TEA, by contrast, preserved a low background and therefore achieved a better signal-to-noise ratio. On the basis of these measurements, the most promising TEA configuration was identified as the addition of a 25~g/L TEA solution directly to the molten emulsion gel at $40\text{-}45~^\circ\mathrm{C}$. 

\subsection{Sensitivity to carbon ions}
\label{HIMACtest} 

\begin{figure}[htbp]
    \begin{center}
        \includegraphics[width=0.5\columnwidth, trim={1cm 1cm 0 0}, clip]{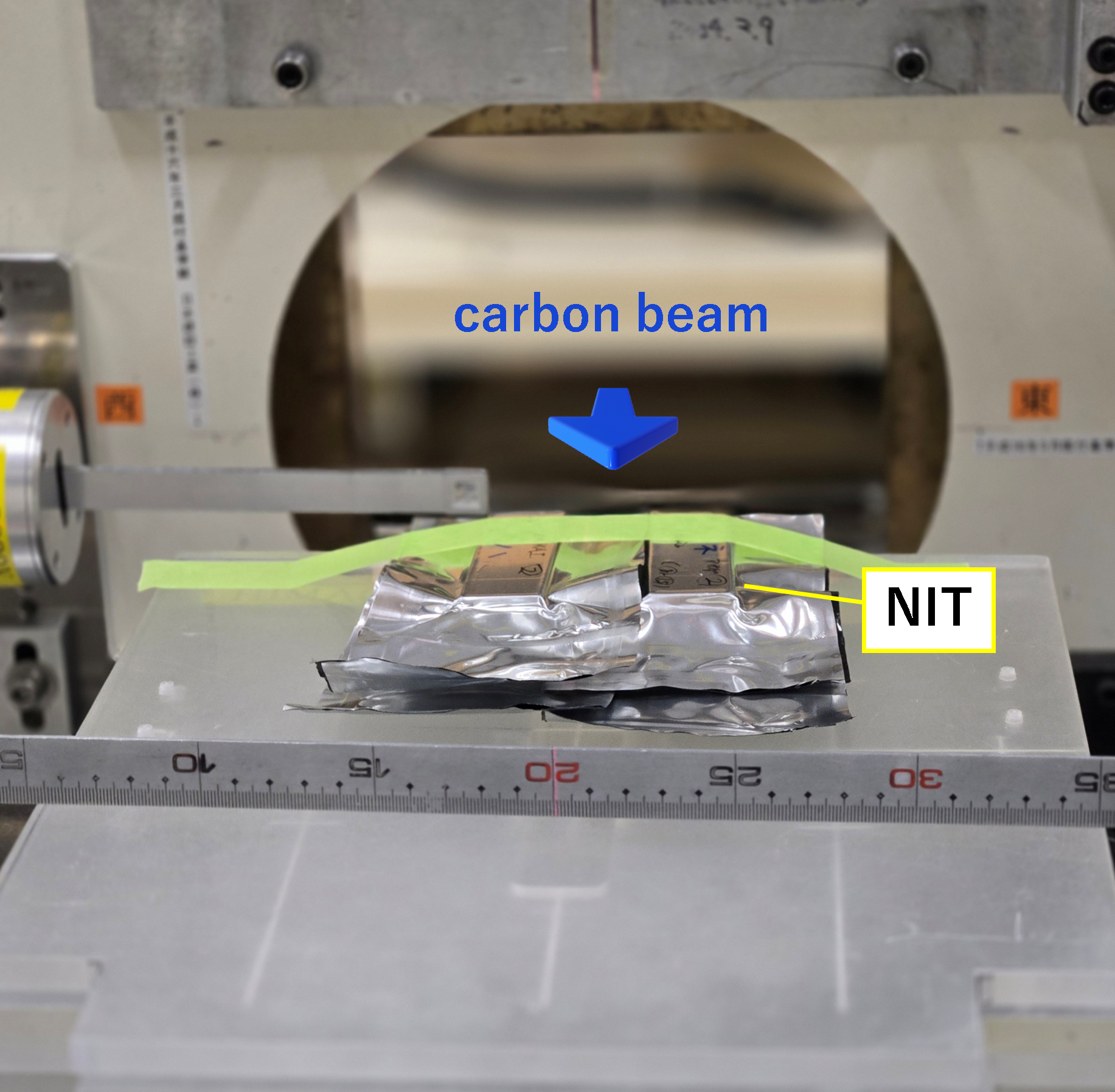}
        \caption{Exposure of NIT slide glasses to 290 MeV/u carbon ions at HIMAC. The direction of the beam was parallel to the film surface. }
    	\label{fig:himacexposure}
    \end{center}
\end{figure}
\begin{figure}[htbp]
    \begin{center}
        \includegraphics[width = \linewidth]{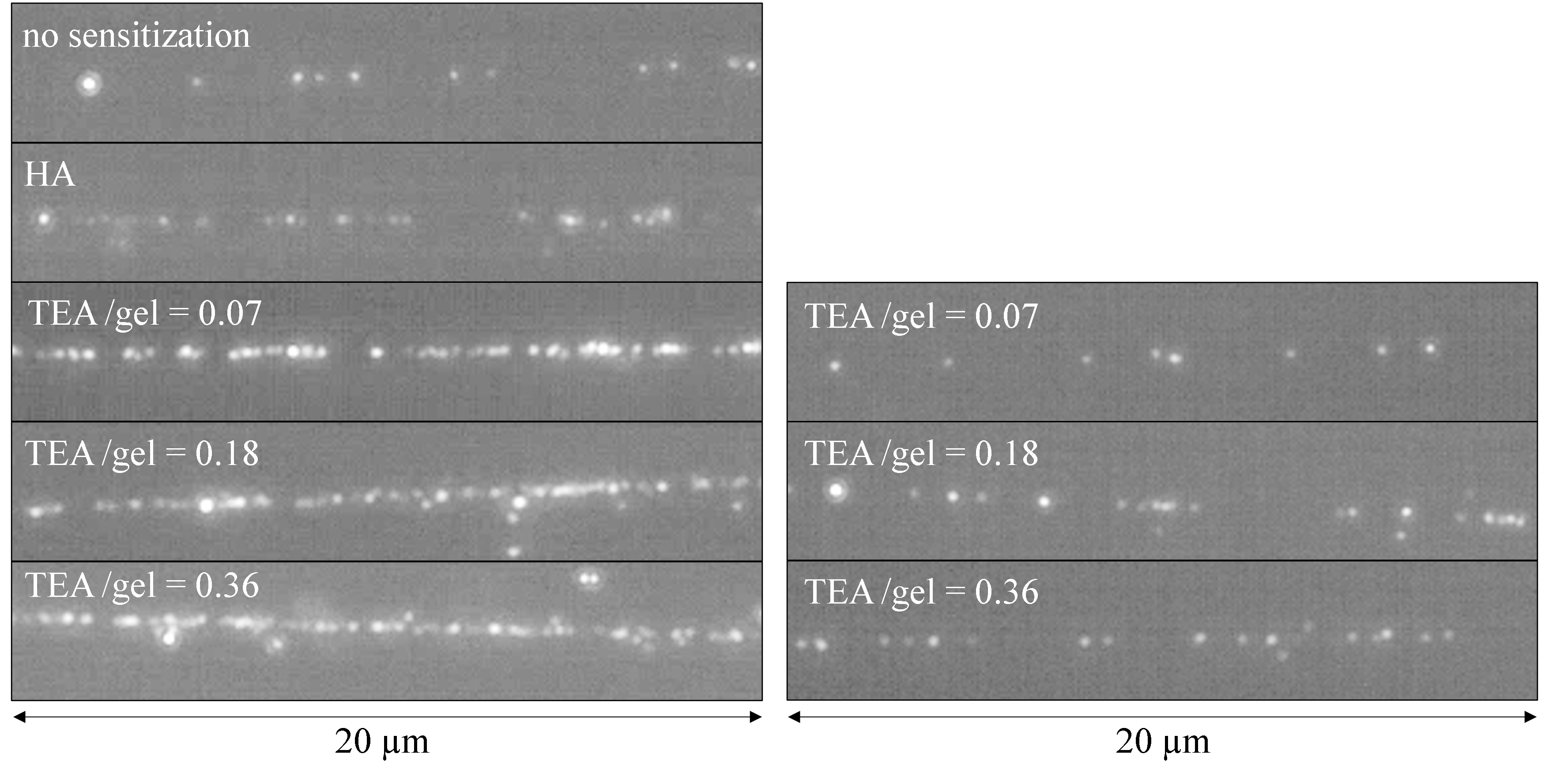}
        \caption{Carbon tracks recorded in NIT under different sensitization conditions: only sensitized NIT (left), and sensitization combined with density dilution such that the crystal density reached $2.0~\si{\per\micro\meter}$ (right). Crystal density dilution reduces the overlap of developed silver grains, improving their optical separation and enabling a better analysis of the effect of TEA.} 
    \label{fig:tea_tracks_differentdensities}
    \end{center}
\end{figure}
\begin{figure}[htbp]
    \begin{center}
        \begin{tabular}{c}
            \begin{minipage}{0.47\linewidth}
                \begin{center}
                    \includegraphics[clip,width=\linewidth]{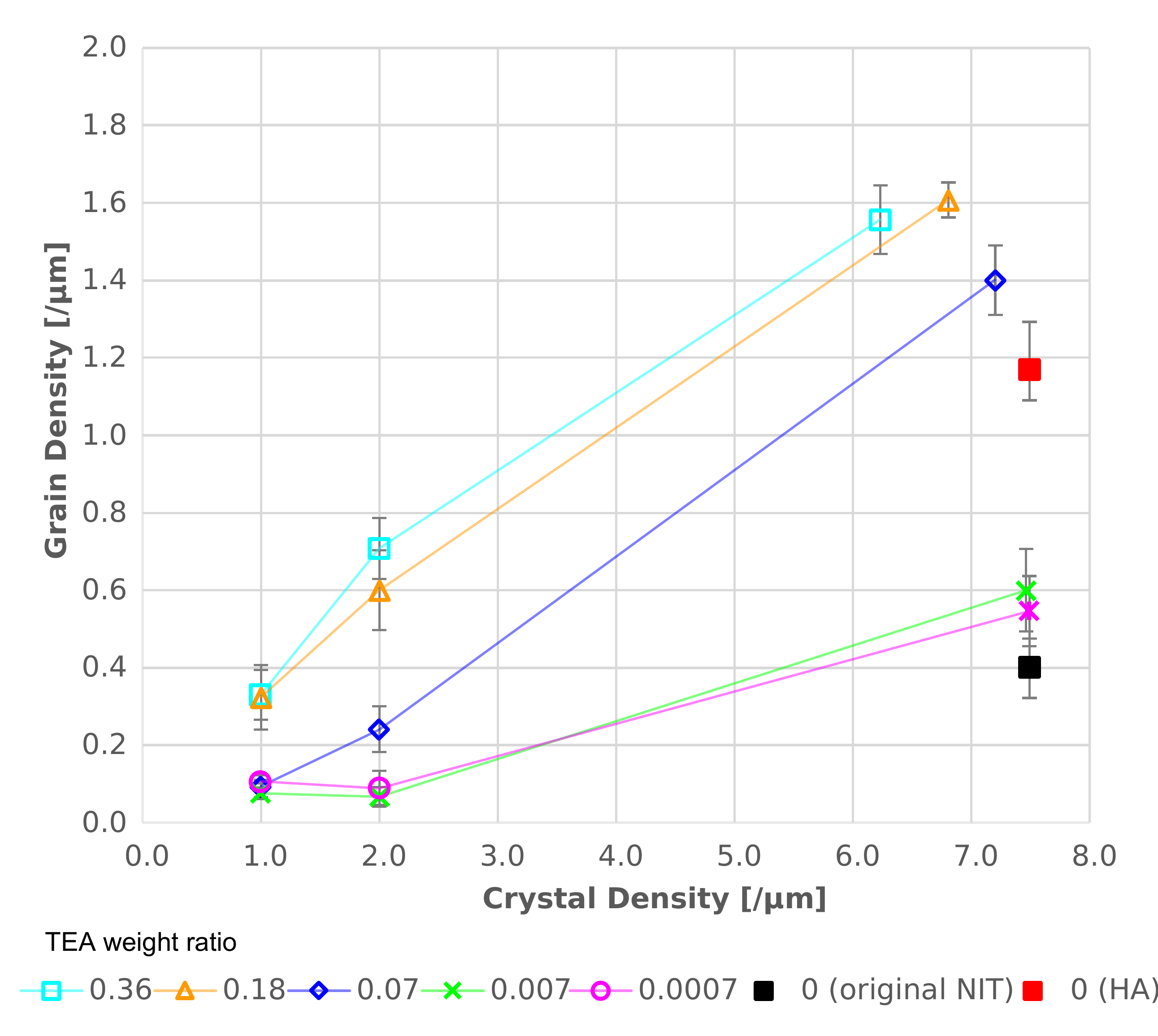}
                    \caption{\label{fig:himac_TEAGDvsCD} Measured grain density in NIT as a function of the calculated crystal density. The legend indicates the weight ratio of TEA to gelatin in dried NIT for the TEA-sensitized samples.
                    }
                \end{center}
            \end{minipage}
              \hspace{0.02\linewidth}
            \begin{minipage}{0.47\linewidth}
                \begin{center}
                    \includegraphics[clip, width=\linewidth]{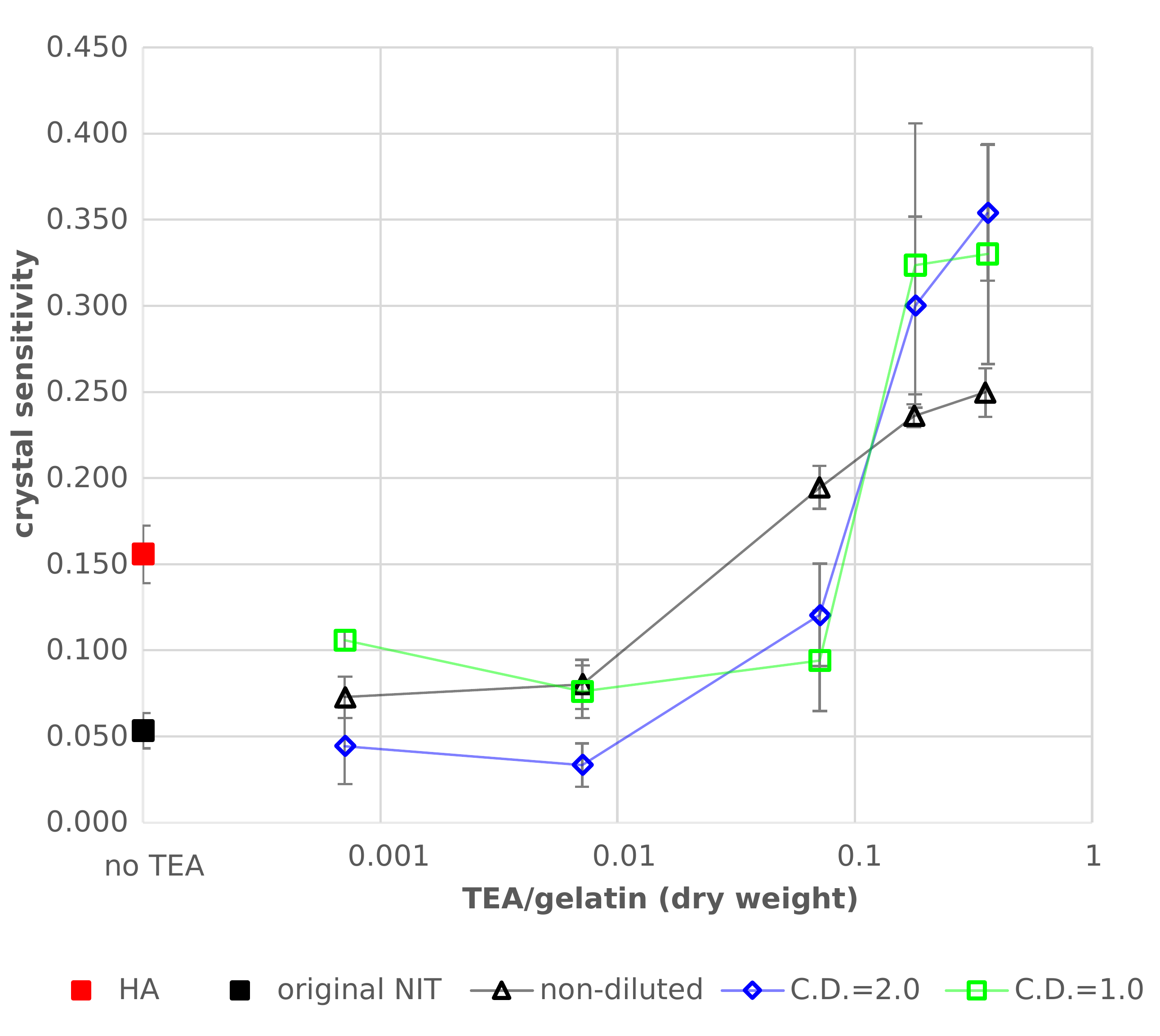}
                    \caption{\label{fig:himac_TEAcrystalsensitivity} Estimated crystal sensitivity per silver halide crystal in NIT as a function of the weight ratio of TEA to gelatin in dried NIT. Each legend indicates the NIT dilution condition.
                    }
                \end{center}
            \end{minipage}
        \end{tabular}
    \end{center}
\end{figure}

To further investigate TEA sensitization and identify the optimal operating conditions, a second study was carried out at HIMAC (Heavy Ion Medical Accelerator in Chiba, Japan \cite{kodaira2024space}), operated by the National Institutes for Quantum Science and Technology. A 290~MeV/u carbon-ion beam was used as a controlled source of highly ionizing particles. For these particles, the estimated stopping power in NIT is approximately \SI{250}{MeV/cm}. 
NIT samples were prepared by dissolving the emulsion at \SI{40}{\celsius}, adding TEA at fixed ratios with respect to the gelatin content and pouring the mixture onto glass slides. For comparison, HA-sensitized samples were also prepared according to the standard procedure described in Sec.~\ref{sec:HAsensitization}, by adding a 0.5\% sodium sulfite solution corresponding to 17.6\% of the molten emulsion weight. 
The carbon beam was incident parallel to the film surface, as shown in Fig.~\ref{fig:himacexposure}. After exposure, the films were developed with the MAA developer and analyzed by optical microscopy. 
Fig.~\ref{fig:tea_tracks_differentdensities} shows horizontal carbon ion tracks in the developed NIT films. During chemical processing, undeveloped silver halide crystals and residual TEA are removed, resulting in vertical shrinkage. Because the tracks were oriented horizontally, this shrinkage does not affect the measured grain density along the track direction and thus does not affect the measurement.
A progressive increase in sensitivity is observed from unsensitized to HA-sensitized and finally to TEA-sensitized samples. However, at high TEA concentrations, individual silver grains become increasingly difficult to resolve due to excessive grain density. The original crystal density, defined as the number of AgBr(I) crystals per unit length, was estimated from the crystal size, the emulsion density and the amounts of added gelatin and TEA. It decreases from \SI{7.5}{\per\micro\meter} for the unsensitized sample to \SI{6.2}{\per\micro\meter} for the sample with the highest TEA-to-gelatin ratio of 0.36, indicating that many developed grains become optically indistinguishable under these conditions.  
To mitigate this effect, diluted NIT samples were prepared with adjusted crystal densities of 1.0 and \SI{2.0}{\per\micro\meter}, as shown in the right image of Fig.~\ref{fig:tea_tracks_differentdensities}. In these samples, an increase in sensitivity was clearly observed with increasing TEA content, while maintaining sufficient spatial separation between developed silver grains for optical identification.
Figure~\ref{fig:himac_TEAGDvsCD} shows the measured grain density as a function of crystal density for different TEA concentrations. Tracks were analyzed within a single focal layer with a depth of \SI{0.3}{\micro\meter}, defined by the microscope focal depth. If the sensitization effect were uniform, the grain density would scale proportionally with the crystal density. This behavior is observed at low crystal density, while deviations appear at higher values, confirming that grain overlap limits optical separability. 
The crystal sensitivity, defined as the ratio between grain density and crystal density, is shown in Fig.~\ref{fig:himac_TEAcrystalsensitivity}. For samples with crystal densities of 1.0 and 2.0~\si{\per\micro\meter}, the crystal sensitivity follows a consistent trend as a function of TEA concentration, indicating that the grains are sufficiently separated. Under these conditions, the TEA‑to‑gelatin ratio reaches a practical upper limit of 0.18. Beyond this value, further addition of TEA degrades the dispersion of silver halide crystals and may adversely affect the mechanical stability of the emulsion layer. Based on these results, a TEA‑to‑gelatin ratio of 0.18 was identified as the optimal configuration. In practical terms, this corresponds to adding a \SI{2.5}{\percent} TEA solution to the NIT gel dissolved at \SI{40}{\celsius}, in an amount equal to \SI{17.6}{\percent} of the emulsion weight.
    
\section{Advancements in NIT development}
    
\begin{figure}[h!]
\centering
\includegraphics[width = 0.8\linewidth]{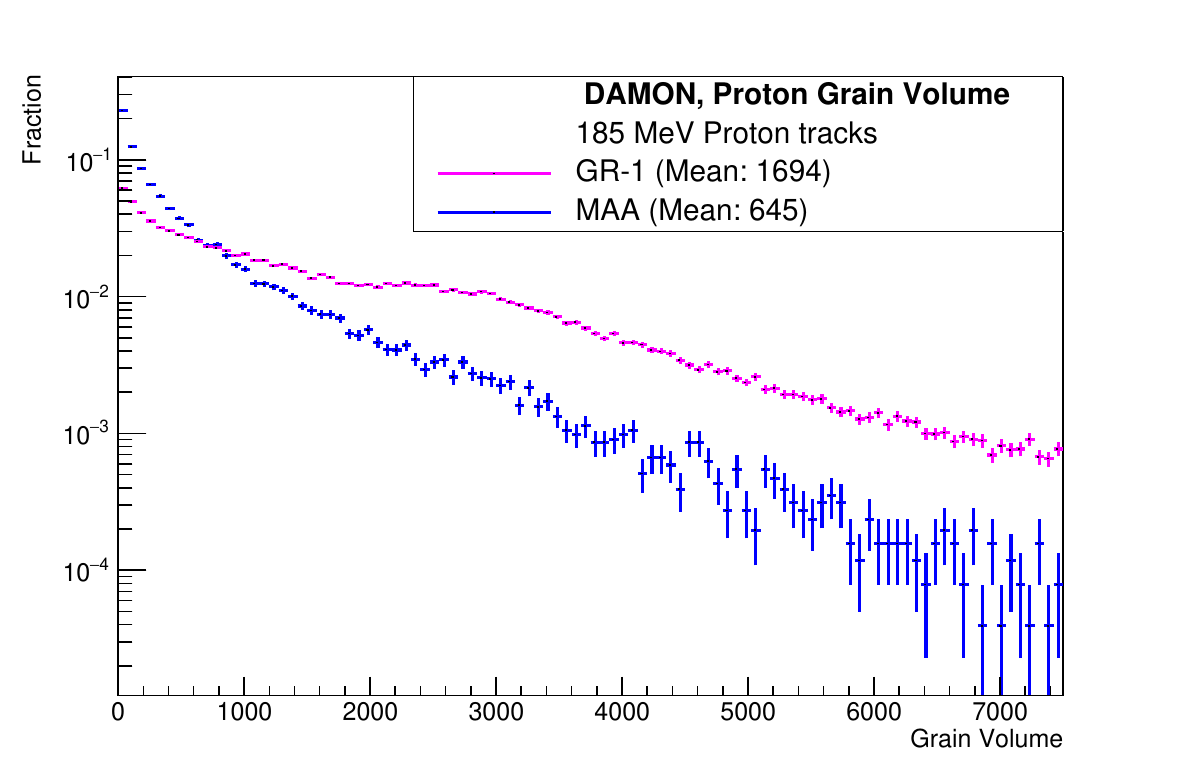}
\caption{Grain volume distributions for 185 MeV horizontal proton tracks recorded in 70 nm NIT films developed with GR-1 (fuchsia) and MAA (blue). The grain volume was obtained by summing the pixel intensities of each reconstructed grain and is expressed in arbitrary units (a.u.). The GR-1 distribution is shifted toward larger grain volumes, indicating the formation of larger and brighter silver grains.} 
\label{fig:grain_volume_comparison_p200}
\end{figure}

In addition to optimizing the sensitization process, alternative development techniques capable of enhancing grain contrast and visibility were investigated. In particular, a physical development approach was considered, in which silver halide crystals are partially dissolved during development, providing silver ions that contribute to the growth of metallic silver grains. This mechanism results in larger developed grains and increased optical contrast. 
In this context, the GR-1 developer, widely used in photographic chemistry and recently introduced in nuclear emulsion studies, was selected. The GRAINE experiment~\cite{takahashi2015graine} has successfully employed GR-1 to enhance the contrast of developed grains in emulsion films~\cite{yamamoto2023Reversaldevelopment}. In this work, GR-1 was applied under conditions comparable to those used for MAA development, namely at a temperature of \SI{5}{\celsius} and for a development time of 10~min. 
To evaluate the performance of the GR-1 developer, TEA-sensitized NIT samples were irradiated with proton beams at the Nagoya Proton Therapy Center \cite{toshito2016proton}. After exposure, the samples were developed using both the standard MAA developer and GR-1, and subsequently analyzed by optical microscopy at the University of Naples Federico II. 
\begin{figure}[h!]
\centering
\includegraphics[width = 0.9\linewidth]{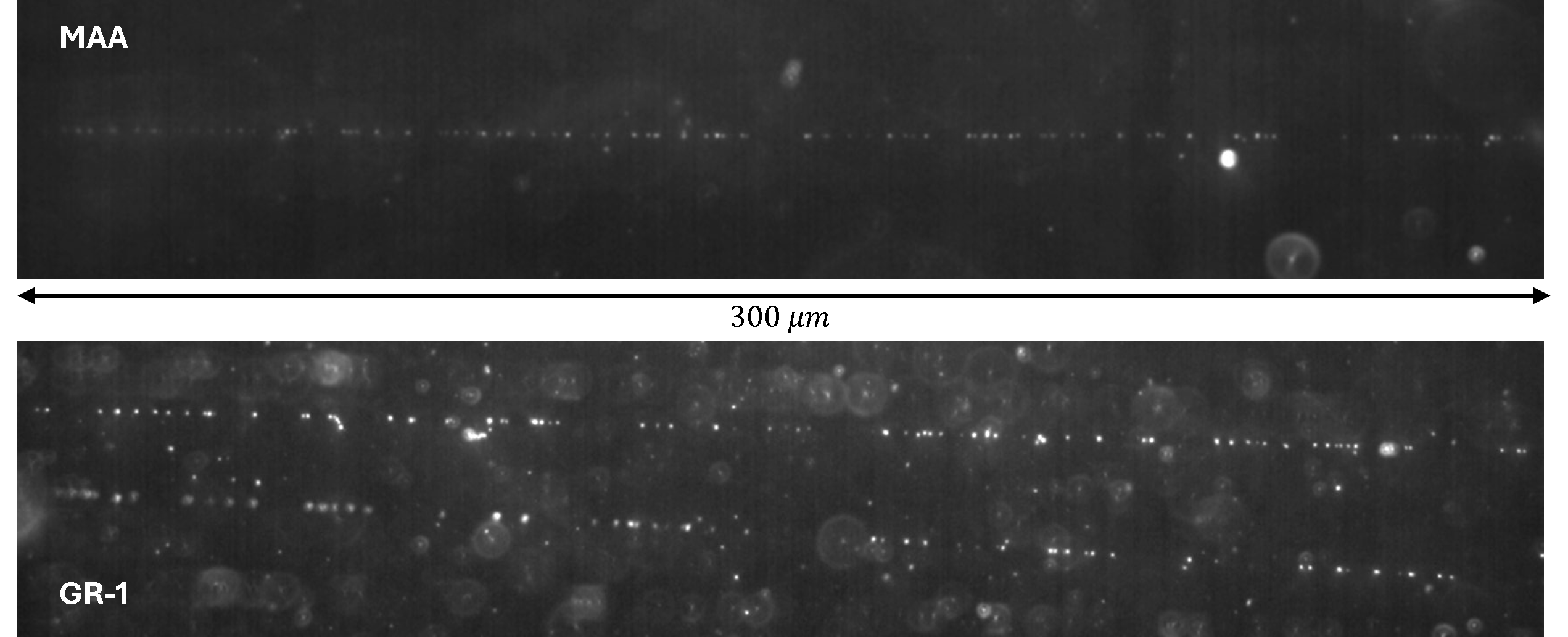}
\caption{Representative horizontal 70 MeV proton tracks recorded in 70 nm NIT films developed with MAA (top) and GR-1 (bottom). Images were acquired using the same microscope under identical illumination conditions. The GR-1 developer produces systematically brighter silver grains than MAA, improving track visibility. Short delta rays emitted from the primary proton track are also visible.}
\label{fig:visual_comp}
\end{figure}
Figure~\ref{fig:visual_comp} shows representative horizontal proton tracks recorded in 70~nm NIT films. To ensure a consistent comparison, images were acquired using the same microscope under identical illumination conditions. A clear increase in the brightness of developed grains is observed in the GR-1 samples compared to those developed with MAA. Short delta rays originating from the primary track are also visible, suggesting possible applications of NIT in microdosimetry.  
The integrated brightness of each grain was evaluated by summing the pixel values of all of its pixels. This quantity, referred to as the grain ``volume'' and expressed in arbitrary units (a.u.), provides a measure of the optical signal associated with each developed grain. The corresponding distributions are shown in Fig.~\ref{fig:grain_volume_comparison_p200}.
The GR-1 sample exhibits a systematic shift toward larger grain volumes, indicating an overall increase in both grain size and optical contrast. In MAA-developed films, only about 9\% of grains have volumes exceeding 2000~a.u., whereas this fraction increases to approximately 28.5\% for GR-1. This result demonstrates that GR-1 produces larger and brighter silver grains, thereby improving the visibility of proton tracks, especially at higher energies.

\section{Experimental validation with protons}

\subsection{Exposure configuration}

\begin{figure}[h!]
\centering
\includegraphics[width = 0.9\linewidth]{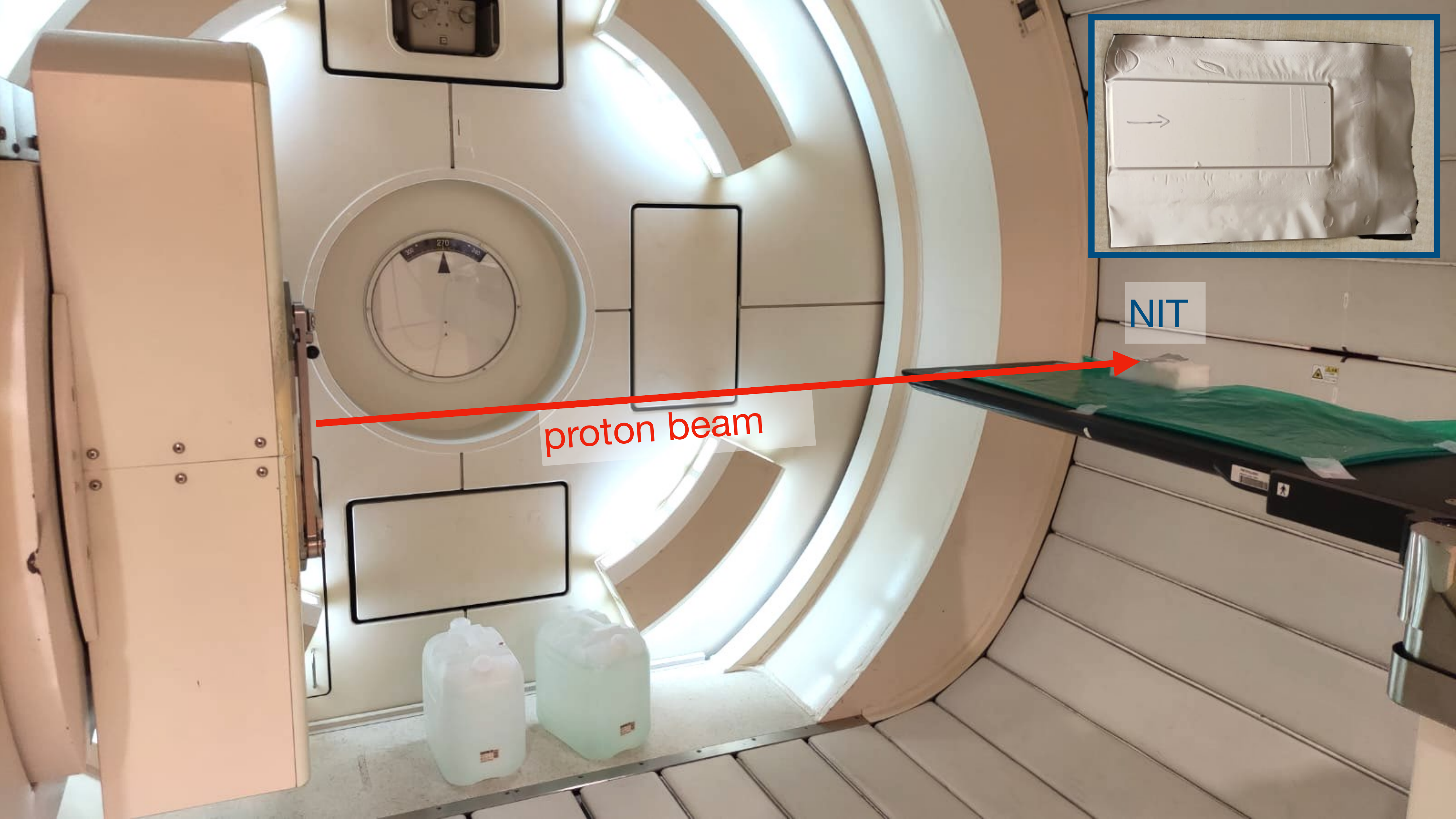}
\caption{Experimental setup for the horizontal irradiation of NIT samples with a clinical proton beam at the Nagoya Proton Therapy Center. The inset shows the samples placed inside a vacuum sealed light tight bag.}
\label{fig:nagoya_exposure_1}
\end{figure}

The response of TEA-sensitized NIT developed with GR-1 was investigated through measurements performed at the Nagoya Proton Therapy Center. TEA-sensitized emulsion gel was poured onto glass slides, which were then exposed to proton beams in a horizontal configuration, as shown in Fig.~\ref{fig:nagoya_exposure_1}. 
Proton energies at the isocenter were set to 200, 140, and 70~MeV, covering the full range of therapeutic beams. The horizontal geometry allowed a single proton track to be sampled continuously at different depths within the emulsion layer, corresponding to different kinetic energies along the track. Intermediate energy values were estimated using a \textsc{Geant4}-based Monte Carlo simulation. The main observable used to characterize the detector response was the grain density, defined as the number of developed grains per unit length along the reconstructed track. For each energy point, the grain density was averaged over ten randomly selected tracks to reduce statistical fluctuations.

\subsection{Proton grain density calibration}

\begin{figure}[h!]
\centering
\includegraphics[width = 0.95\linewidth]{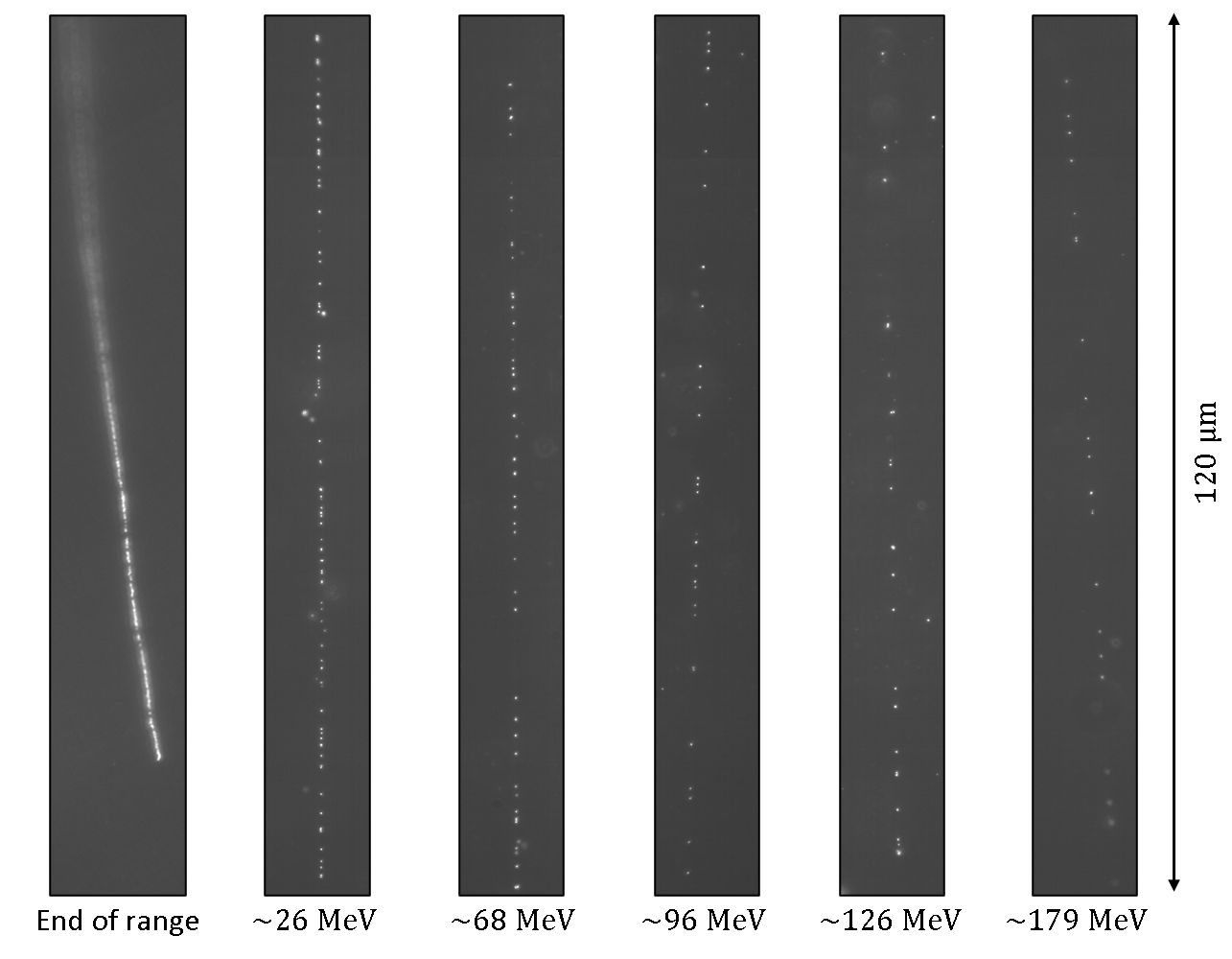}
\caption{Sections of proton tracks recorded at different depths in TEA-sensitized NIT. The residual proton energy is indicated for each image. This was estimated by simulating with \textsc{Geant4} the proton transport from its initial energy (200, 140, or 70 MeV) through the corresponding thickness of emulsion traversed before reaching the imaged section. As the proton energy increases, the grain density decreases, resulting in reduced track visibility. }
\label{fig:proton_grain_density_nit}
\end{figure}

\begin{figure}[htbp]
    \centering
    \begin{minipage}{0.50\linewidth}
        \centering
        \includegraphics[width=\linewidth]{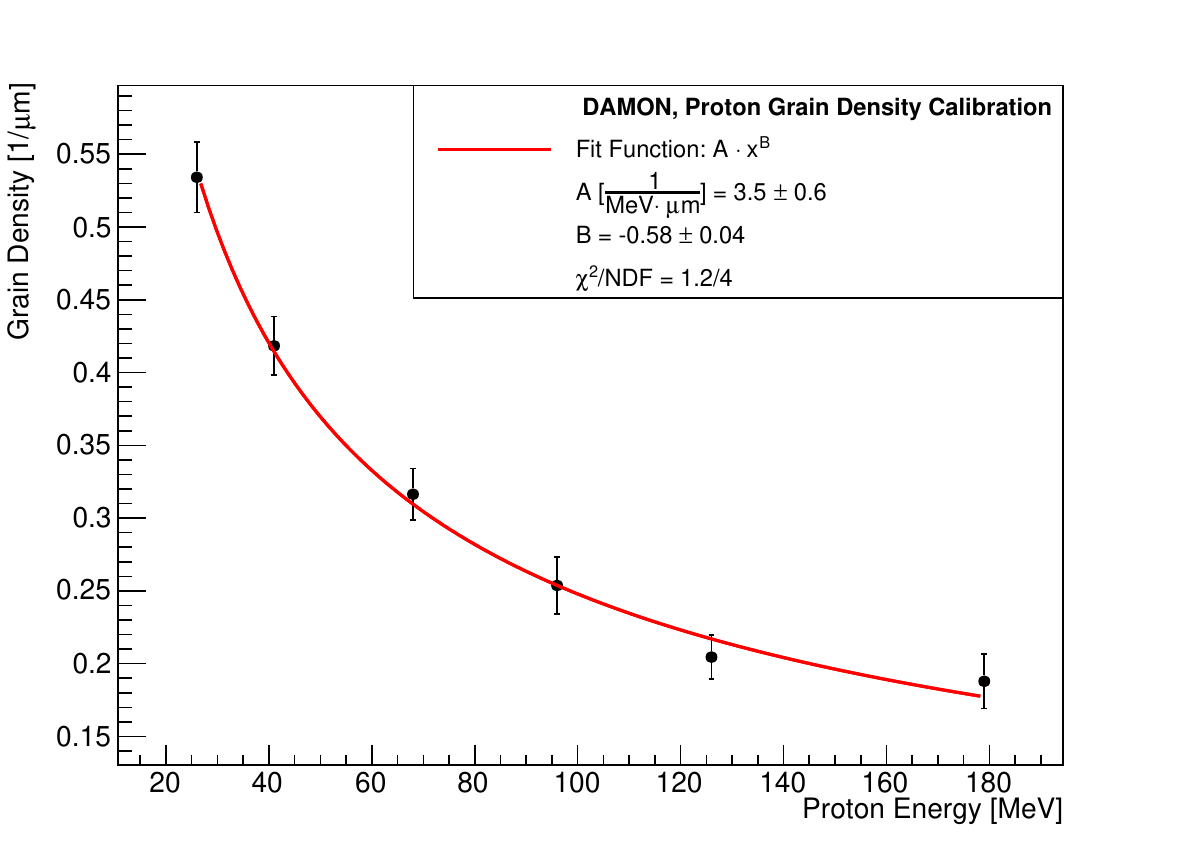}
    \end{minipage}\hfill
    \begin{minipage}{0.50\linewidth}
        \centering
        \includegraphics[width=\linewidth]{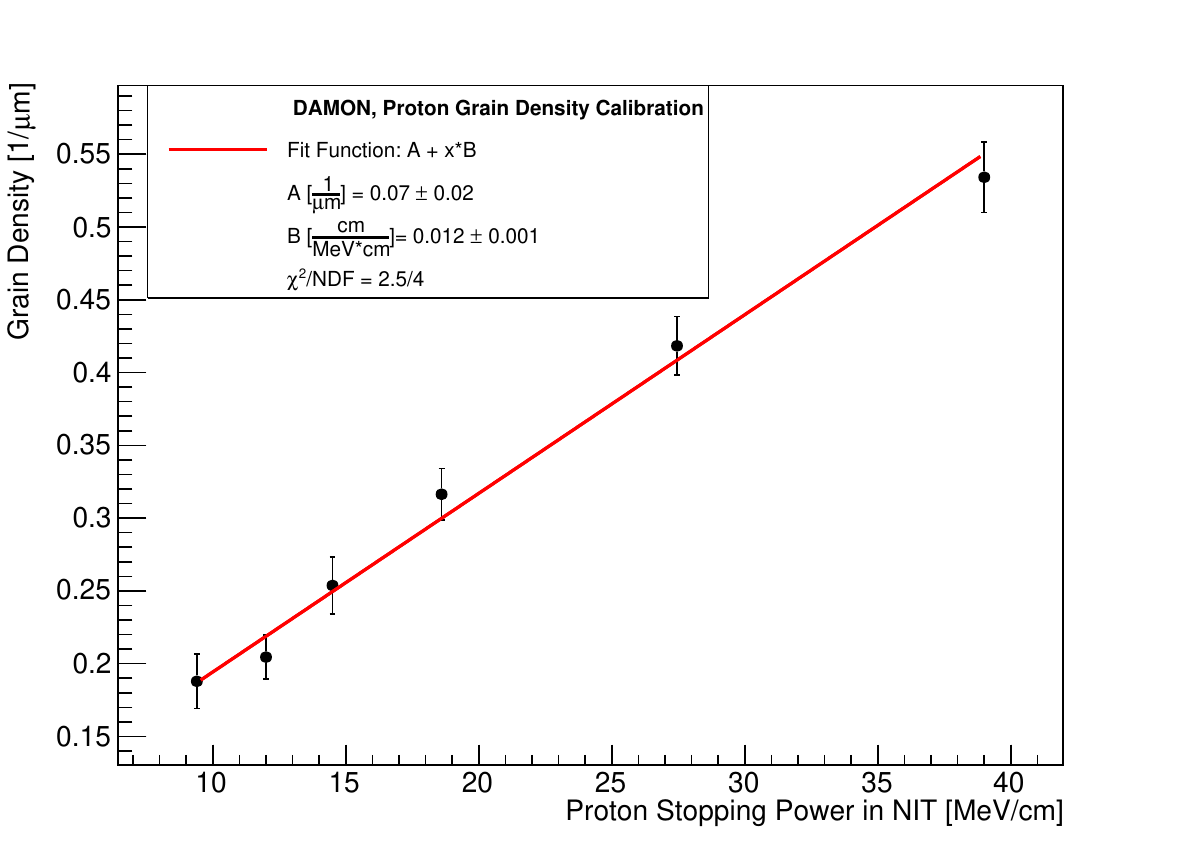}
    \end{minipage}
    \caption{Measured grain density in TEA-sensitized NIT as a function of (left) proton kinetic energy and (right) proton stopping power in NIT, estimated with SRIM \cite{ziegler2010srim}. }
    \label{fig:proton_grain_density}
\end{figure}

Representative proton tracks recorded in TEA-sensitized NIT are shown in Fig.~\ref{fig:proton_grain_density_nit}. As the proton energy increases, the grain density decreases, reflecting the reduction in stopping power. 
The grain density as a function of proton kinetic energy is shown in Fig.~\ref{fig:proton_grain_density}. The uncertainty in the grain density was calculated assuming Poisson statistics for the grain counts and performing a weighted average over the different tracks considered. At higher energies, the grain density decreases below 0.2~grains/\si{\micro\meter}, making track reconstruction increasingly challenging. Within the present scanning and reconstruction conditions, proton tracks could be reconstructed up to about 100 MeV, marking a significant improvement over HA-sensitized NIT. Figure~\ref{fig:proton_grain_density} shows a linear correlation between grain density and proton stopping power in NIT, as evaluated with SRIM~\cite{ziegler2010srim}. Within the investigated energy range, this behaviour indicates that TEA-sensitized NIT can provide an indirect estimate of proton kinetic energy through grain-density measurements. Repeated measurements performed on independent samples yielded consistent results, confirming the reproducibility of the detector response.

\color{black}
\section{Conclusions}

In this work, the sensitivity of Nano Imaging Trackers (NIT) was improved through the use of alternative chemical sensitizers (gold-plus-sulfur and triethanolamine) and the GR-1 developer. Measurements at LNGS showed that both Au-S and TEA enhance latent-image formation with respect to the standard HA sensitization. While Au-S provided the largest sensitivity increase, it also produced a significant increase in fog density. TEA sensitization instead achieved a more favorable balance between sensitivity and background, resulting in an improved signal-to-noise ratio.
Measurements with carbon ions at HIMAC identified an optimal TEA-to-gelatin dry-weight ratio of 0.18, beyond which the gain in sensitivity does not compensate for the potential degradation of the emulsion's mechanical properties. In addition, the GR-1 developer was successfully extended to NIT with 70~nm crystals, significantly increasing the size and brightness of developed grains.
Measurements at the Nagoya Proton Therapy Center demonstrated a linear correlation between grain density and stopping power, indicating that NIT can provide an indirect measurement of proton kinetic energy. The optimized sensitization and development procedure enabled the reliable reconstruction of proton tracks up to approximately 100~MeV, representing a substantial improvement over the previous HA-based formulation. 
These developments significantly enhance the performance of NIT detectors for proton therapy applications, enabling a more complete characterization of secondary radiation from target-fragmentation events and supporting future applications in microdosimetry.

\acknowledgments

This work was supported by the European Union -- NextGenerationEU, Mission 4, Component 1 (CUP H53D23001090006), by JSPS KAKENHI Grant Nos. JP23KK0058 and JP24H02241, and by the Joint Research Project of the Heavy-Ion Medical Accelerator in Chiba (HIMAC) at the National Institutes for Quantum Science and Technology (QST), Project Nos. H212 and HH006. The authors also thank Dr. Toshiyuki Toshito of the Nagoya Proton Therapy Center and Prof. Ken'ichi Kuge of Chiba University for his valuable suggestion of TEA sensitization.



\bibliographystyle{JHEP}
\bibliography{biblio.bib}


\end{document}